# Enhancing interfacial thermal transport by nanostructures: Monte Carlo simulations with ab initio phonon properties

Wenzhu Luo [a†], Neng Wang [a†], Wenlei Lian [bc], Ershuai Yin [a*], Qiang Li [a*]

[a] MIIT Key Laboratory of Thermal Control of Electronic Equipment, School of Energy and Power Engineering, Nanjing University of Science & Technology, Nanjing, Jiangsu 210094, China

[b] College of Energy and Power Engineering, Nanjing University of Aeronautics and Astronautics, Nanjing, Jiangsu 210016, China

[c] Key Laboratory of Thermal Management and Energy Utilization of Aviation Vehicles, Ministry of Industry and Information Technology, Nanjing, Jiangsu 210016, China

**Abstract:** Recent experiments have indicated that employing nanostructures can enhance interfacial heat transport, but the mechanism by which different structural morphologies and dimensions contribute to the full-spectrum phonon interfacial transport remains unclear. In this paper, a multiscale method to study the thermal transfer at nanostructured interfaces is developed by combining density functional calculation, Monte Carlo simulation, and diffuse mismatch method. The changes in the transport paths and contributions to thermal conductance of different frequency phonons caused by changes in nanostructure morphology and size are investigated. The results show that, compared to the triangular and trapezoidal nanostructures, the rectangular nanostructures are more beneficial in enhancing the probability of the reflected phonons encountering the interface, and thus the phonon interfacial transmittance. The nanostructure makes the interfacial heat flow extremely heterogeneous, with significant transverse heat flow occurring at the sidewalls, resulting in a new thermal conduction pathway. The phenomena of multiple reflections and double transmission together lead to the existence of the optimal dimension that maximizes the nanostructures enhancement effect on interfacial heat transfer. The optimal nanostructure width is 100 nm when the height is 100 nm and the maximum interfacial thermal conductance enhancement ratio is 1.31. These results can guide the design of heat transfer enhancement structures at the interface of the actual high-power chips.

[†] These authors contributed equally to this work.

[*] Corresponding authors. E-mail address: yes@njust.edu.cn (E. Yin), liqiang@njust.edu.cn (Q. Li)

# 1 Introduction

With the gradual increase in the power density of electronic devices, effective thermal management has become the key to ensuring efficient and stable operation of the devices [1]. The use of various types of high thermal conductivity material substrates (e.g., diamond) [2] has made the interfacial thermal resistance caused by the lattice mismatch [3], defects [4], and other factors progressively become the governing thermal resistance inside the chip [5]. For example, test results indicated that for the AlGaN/GaN-on diamond transistor, the temperature rise induced by the interfacial thermal resistance accounted for more than 40% of the total temperature rise of the device [6]. Therefore, enhancing interfacial thermal conductance (ITC) has become a hot research topic in the field of electronic device thermal management [7,8].

Recently, the nanostructured interface method has been proposed and experimentally proven to be effective in improving interfacial heat transfer [9–11]. For example, Cheng et al. [12] prepared the Si/diamond interface containing two-dimensional rectangular nanostructures. They found that employing nanostructures increased the ITC by up to 65%, from 63.7 $MWm^{-2}K^{-1}$ to 105 $MWm^{-2}K^{-1}$, compared to the flat interface. Lee et al. [13] fabricated Al/Si interfaces containing three-dimensional arrays of nanostructures, which enhanced the ITC by 88% compared to planar interfaces. Park et al. [14] measured the equivalent thermal conductivity of the Al/$SiO_2$ interface containing the rectangular nanostructure array. Compared to the planar interface, using the nanostructured interface achieves up to more than 2-fold increase in the interfacial equivalent thermal conductivity. These experimental studies have all demonstrated that the nanostructures significantly increase the actual contact area of the two materials, which is the primary reason for the improved interfacial heat transfer characteristics. However, the experimentally obtained ITC enhancement ratios of the nanostructured interfaces are all lower than the theoretical results that only take into account the role of area enhancement [13]. This implies the existence of other effects weakening the ITC enhancement of nanostructures, which together with the contact area boost determine the nanostructured interface ITC.

Some theoretical studies have been conducted to reveal the mechanism of ITC reinforcement by nanostructured interfaces [15–17]. For example, Qi et al. [18] studied the effect of a rectangular nanopillar array on the thermal resistance of the AlN/diamond interface using molecular dynamics simulations and achieved a 28% reduction in the interfacial thermal resistance compared to the planar surface. They attributed this enhancement to the alteration of the mid/high-frequency phonons vibrational density of states (VDOS) resulting from the nanostructured interface. Xu et al. [19] also used molecular dynamics simulations to investigate the structural size and arrangement of rectangular nanopillar arrays on Si/4H-SiC interfaces and achieved up to 11% ITC enhancement. They suggested that the ITC-enhancing effect at nanostructured interfaces was the result of a combination of phonon boundary scattering and interfacial phonon transport channels. Hua et al. [20] studied the effect of two-dimensional rectangular nanostructures on ITC using a grays-media approximation phonon Monte Carlo method. They concluded that the multiple reflections of phonons at the nanostructured interface are the main reason for boosting the ITC, which increased the probability of phonon transmission and created a thermally conductive pathway. Zhao et al. [21] similarly investigated the effect of rectangular structure size on interfacial ITC using the Monte Carlo method and found that the interfacial thermal conductance was maximum when the contact area ratio was 2.0. They attributed this to a combination of contact area and interfacial backscattering. We have studied the effect of large-size nanostructure morphology and size on the ITC of the GaN/diamond interface using the lattice Boltzmann method and achieved an ITC enhancement of 159.6% [22]. It was found that the nanostructure morphology significantly affects the ITC, but the behavior of the interface phonons cannot be accurately described due to adopting the grays-media approximation, which may cause the results to deviate from the actual values [23]. Therefore, the scattering and transport of phonons for all frequencies and modes should be comprehensively considered to further explore the effects of morphology and size on thermal conductance at nanostructured interfaces. This is important because the interfacial structures prepared are difficult to guarantee as ideal rectangular nanostructures [13]. According to the diffuse mismatch method (DMM), the phonon interface transmittance is correlated with the incident angle [24,25]. This implies that the different inclinations of the nanostructures' sidewalls will affect the phonon transmittance at the interface and hence the interfacial thermal conductance. Since phonons with different frequencies have distinct properties and transmittances [26,27], they may benefit differently from the area boost afforded by nanostructures. Moreover, changes in nanostructure size can significantly alter the interfacial transport of phonons with large mean free paths, thus affecting their contribution to the interfacial thermal conductance. It is necessary to study the changes in the transport paths and contributions to the thermal conductance of different frequency phonons caused by changes in nanostructure morphology and size. This understanding is crucial for clarifying the mechanism of enhanced interfacial thermal transport by nanostructures and maximizing interfacial thermal conductance.

In this paper, a multiscale method to study the thermal transfer at nanostructured interfaces is developed by combining density functional calculation, Monte Carlo simulation, and diffuse mismatch method. The changes in the transport paths and contributions to thermal conductance of different frequency phonons caused by changes in nanostructure morphology and size are investigated. The optimal nanostructure morphology and size are determined, and the mechanism of enhancing interfacial phonon transport by nanostructures is revealed. This study can provide guidance for the design of heat transfer enhancement structures at the interface of the actual high-power chips.

## 2 Methodology

Figure 1 shows the schematic diagram of the nanostructured and planar interfaces. The computational domain contains two materials, Material 1 and Material 2. This paper takes the Si/Ge interface as an example to study the heat transfer at nanostructured interfaces. Therefore, Material 1 is Si, and Material 2 is Ge. The Si/Ge nanostructured interface is chosen because the interfacial heat transfer of planar structures has been widely studied as a benchmark. For nanostructured interfaces, different trapezoidal structures in the interfacial region are considered. The height of the nanostructure is $H$, the bottom width is $W$, and the spacing is $S$. The top-to-bottom width ratio of the trapezoidal nanostructure is defined as:

$$\beta = \frac{W_T}{W} \tag{1}$$

where $W_T$ is the top width and $W$ is the bottom width. Since $\beta$ can be varied from 0 to 1, the studied structural morphology contains the triangle, rectangle, and different trapezoids. All nanostructures are two-dimensional, as it is beneficial to analyze the interfacial phonon transport behavior and the contribution to the thermal conductance.

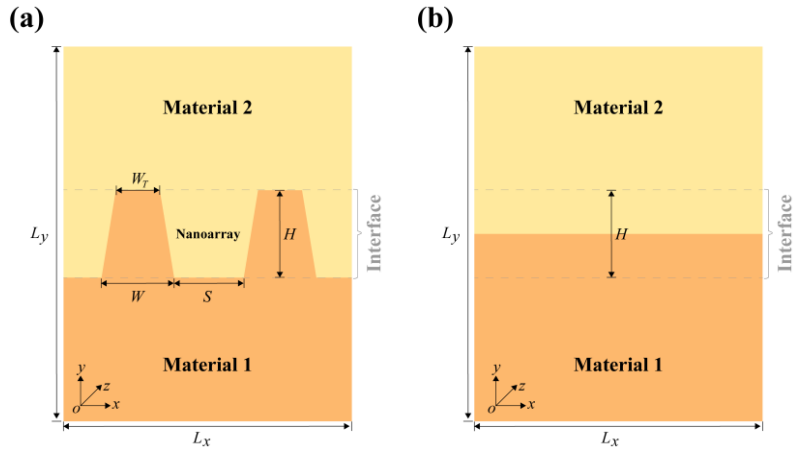

Fig. 1. Schematic diagram of the (a) nanostructured and (b) planar interfaces.

The phonon transport in the computational domain is simulated using the variance-reduced Monte Carlo (MC) method [28]. The MC method can be used to calculate steady-state and transient heat transport within complex geometries by solving the linearized Boltzmann transport equation (BTE), and its basic principles have been presented in reference [29]. The accuracy of BTE calculations depends on the input parameters, including the phonons' angular frequency, group velocity, relaxation time, and specific heat capacity. In previous studies, these parameters were often adopted using isotropic phonon scattering obtained from semi-empirical phonon properties[30,31], which may cause the results to deviate from the actual values [32]. The phonon properties of Si and Ge materials studied in this paper are derived from density functional calculation (DFT) calculations. Firstly, the second-order and third-order force constants of both materials are obtained by combining the Vienna Ab initio Simulation Package (VASP) [33,34], PHONOPY [35], and the thirdorder.py package [36]. The alamBTE software package [37] is subsequently used to obtain all the required phonon properties.

In the MC simulation, the bottom of the computational domain is set to 303 K and the top is set to 297 K, resulting in a heat flow from the bottom upwards. Periodic boundary conditions are used for the other four faces on the left and right (x-direction) and front and back (z-direction). When the phonon leaves a periodic boundary, it enters the boundary corresponding to it maintaining the same information. During the calculation, phonons are emitted from the constant temperature boundary and undergo phonon-phonon and phonon-interface scattering until they hit the constant temperature boundary again and disappear. When phonons encounter the interface, the phonons may transmit through the interface or be reflected. The phonon transmittance at the interface is calculated based on the scattering mismatch theory (DMM) [38], expressed as:

$$\tau_{1\to 2}(\omega') = \frac{\Delta V_2 \sum_{\mathbf{k},p} |\mathbf{v}_{g,2}\cdot\mathbf{n}| \delta(\omega'-\omega)}{\Delta V_1 \sum_{\mathbf{k},p} |\mathbf{v}_{g,1}\cdot\mathbf{n}| \delta(\omega'-\omega) + \Delta V_2 \sum_{\mathbf{k},p} |\mathbf{v}_{g,2}\cdot\mathbf{n}| \delta(\omega'-\omega)} \qquad (2)$$

where subscripts 1 and 2 represent material 1 and material 2, $\Delta V$ represents the volumes of the discretized cells corresponding to the Brillouin zones, $\mathbf{v}_g$ is the phonon group velocity, $\mathbf{n}$ is the normal vector pointing from material 1 to material 2, $\delta$ is the Dirac delta function, which is equal to unity when the phonon angular frequency $\omega'$ is equal to $\omega$, and 0 otherwise. $\sum_{\mathbf{k},p}$ denotes summation over the entire Brillouin zone and all phonon modes. The transmittance is correlated with the incidence angle, phonon group velocity, and angular frequency, and can therefore be used to model the effect of different structural morphologies on the thermal transfer at the interface. When calculating the interfacial transmittance, the phonon properties are also taken from the DFT calculations.

The z-direction dimension is kept at 100 nm, while the x-direction dimension is equal to $S + W$, considering only one period. The size in the y-direction is set above 200 nm because, according to our results, the effect of the size on the ITC is negligible when $L_y$ is larger than 200 nm. The total number of phonons is 600,000, which is determined by convergence testing, and continuing to increase the phonon number will not affect the results. In addition, the temperature distribution and heat flux distribution are obtained by setting 40,000 observers (with grids of 20, 100, and 20 in the x, y, and z directions, respectively) to count the temperature and heat flux contributions brought by phonons passing through each observation domain. Subsequently, the thermal conductance of the nanostructured interface can be calculated by the following equation [19,39]:

$$G = \frac{Q}{\Delta T} \qquad (3)$$

where $Q$ is the heat flow through the interface and $\Delta T$ is the average temperature drop at the interface.

The phonon average scattering number is defined to characterize the effect of nanostructures on phonon interfacial transport:

$$N_{ave} = \frac{\sum_{Phonons} N_{interface\ scatting}}{N_{Phonons}} \quad (4)$$

where $N_{interface\ scatting}$ denotes the total interface scattering number of individual phonons and $N_{phonons}$ is the total phonons number. $N_{interface\ scatting}$ is obtained statistically during the model calculations. Once a phonon encounters the interface, its interfacial scattering number is added to one, regardless of whether it is transmitting or not.

The interfacial thermal conductance enhancement ratio $R_G$ and the area gain ratio $R_A$ for nanostructured interfaces versus planar surfaces are defined as:

$$R_G = \frac{G_{Nano}}{G_{plate}} \quad (5)$$

$$R_A = \frac{A_{Nano}}{A_{plate}} \quad (6)$$

where $G_{Nano}$ and $G_{plate}$ are the thermal conductance of the nanostructured and flat interfaces, and $A_{Nano}$ and $A_{plate}$ are the contact areas of material 1 and material 2 in the nanostructured and flat interfaces.

In addition to the above MC method, the heat transfer characteristics of the nanostructured interfaces are calculated using the lattice Boltzmann model (LBM) and compared with the above MC results. This LBM method was established in our previous paper [22] using a lattice structure of D2Q8 to solve the phonon Boltzmann transport equation. The method employs the gray-media approximation, i.e., the heat transfer is calculated in a single phonon mode with a constant transmittance, which implies that all phonons will benefit from the structural area enhancement. Phonons ballistically transmit at the interface, maintaining their original orientation, and the transmittance is independent of the direction of phonon incidence. The parameters required for LBM calculations are also obtained statistically based on first-principles data.

## 3 Results and Discussion

### 3.1 Model verification

The interfacial thermal conductance at the Si/Ge flat interface is calculated using the MC method and compared with the results in the literature, which are illustrated in Figure 2. It can be found that the calculated results are in good agreement with those obtained by different methods in the literature (DMM [40], MC [21] and atomic Green's function, AGF [41]). The ITC of the planar interface is about 218 MW/m$^2$K and is within the range of experimental results measured by the time-domain thermoreflectance (TDTR) method in the literature [42]. The reason for the measured high ITC is the presence of atomic mixing at the actual Si/Ge interface, which alters the phonon spectrum characteristics of the materials [42]. In addition, the effect of computational domain height on ITC is accurately modeled and agrees with the trend reported in the literature [32]. These results confirm the high accuracy of our method for subsequent studies.

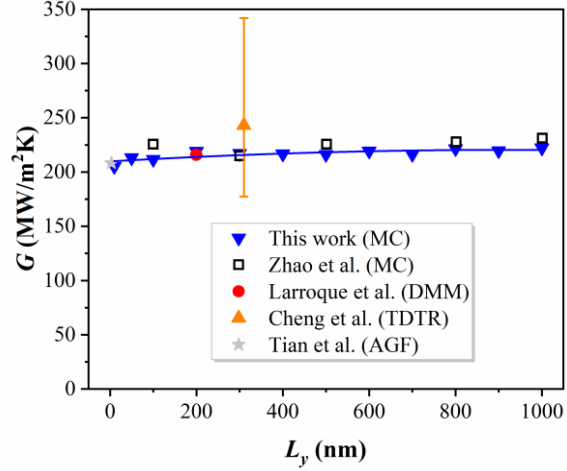

Fig. 2 Theoretical model verification: comparison of thermal conductance at the planar Si/Ge interface for different simulation/test domain heights. The computational results for the MC, DMM, and AGF methods are taken from the literature [21], [40], and [41], while the TDTR test results are from the literature [42].

*3.2 Effects of nanostructure shape*

The effects of nanostructure shape on interfacial thermal conductance are first investigated using both MC and LBM methods. The height, width, and spacing of the nanostructures are controlled at 30 nm, and the top-to-bottom width ratio ($\beta$) of the trapezoidal nanostructure varies from 0 to 1. As $\beta$ increases, the top width of the nanostructure gradually increases, while the shape of the structure changes from triangles to rectangles. According to Figure 3(a), as $\beta$ increases, the interfacial thermal conductance enhancement ratio ($R_G$) calculated by both methods increases, and the growth slows down. However, the $R_G$ calculated by the LBM method for all $\beta$ are superior to those calculated by the MC method. There are two reasons for this result. On the one hand, the employed LBM method neglects the spectral characteristics of phonons. All phonons share the same phonon properties and interface transmittance, which means that all phonons of different frequencies benefit from the expanded contact area. Whereas in practice, low-frequency phonons have higher transmittance, high-frequency phonons have lower transmittance [25,41], implying that high-frequency phonons may benefit more from the enhanced interfacial area. On the other hand, in the LBM method, phonons are ballistically transported at the interface. All phonons undergo specular reflection or transmit into another material keeping the original direction and velocity [20]. This differs from reality because phonons are usually transported quasi-ballistically at interfaces [13]. These two reasons together cause the LBM method using the gray-media approximation to over-predict the heat transfer strengthening effect of nanostructured interfaces. Figure 3(b) gives the variation of the area gain ratio ($R_A$) and the phonon average scattering number ($N_{eav}$) with $\beta$ for the nanostructured interface. Both $R_A$ and $N_{eav}$ increase rapidly as $\beta$ rises. For the triangular structure ($\beta = 0$), $N_{eav}$ is 1.67, and for the rectangular structure ($\beta = 1$), $N_{eav}$ reaches 2.07, while for the flat interface, $N_{eav}$ is 1.03. The results show that nanostructures significantly enhance the probability of phonons encountering the interface. The nanostructures greatly increase the probability of phonons encountering the interface, fully benefiting from the area enhancement. However, the variation of $R_G$ with $\beta$ is inconsistent with $N_{eav}$, implying that other factors collectively govern the strengthening effect of nanostructured interfaces.

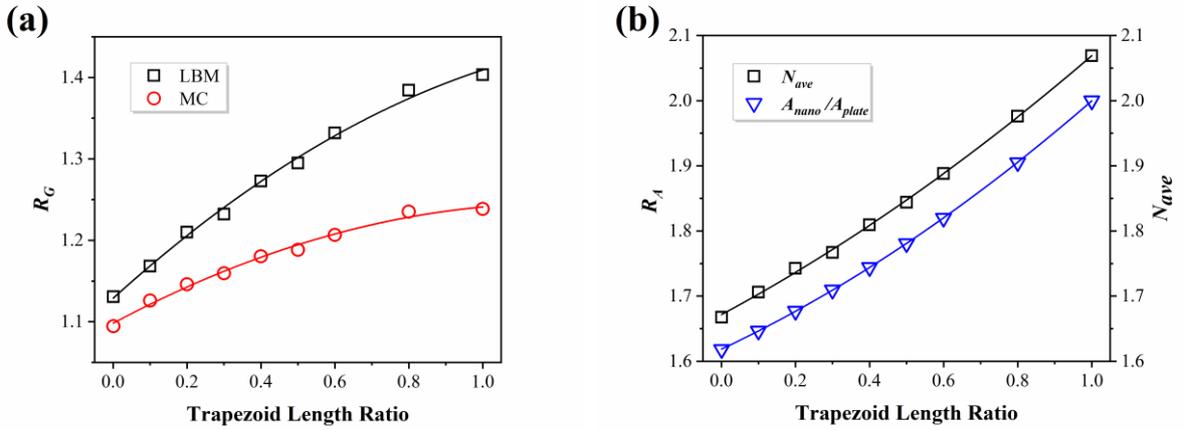

Fig. 3. Effects of nanostructure shape on (a) $R_G$ (b) $R_A$ and $N_{eav}$.

The temperature and heat flow distributions for planar and different shapes of nanostructured interfaces are presented in Figure 4. As can be seen in Figure 4(a-d), the large temperature drops of all the structures occur at the interface. The main difference is that due to the presence of nanostructures, the heat flow, when transported from bottom to top, converges near the nanostructures and is subsequently transported through the side walls of the nanostructures before converging on top of the nanostructures. The existence of nanostructures alters the thermal transfer direction, providing additional paths at the sidewalls. According to Figure 4(e-h), nanostructures cause significant nonuniformity in the interfacial heat flow. Taking a rectangular nanostructure as an example, most of the y-direction heat flux gathers in the interior of the nanostructure and spacer region, while the y-direction heat transfer at the contact interface is low. From the x-direction heat flow in Figure 4(i-l), it can be seen that there is a significant thermal concentration at material 1 near the nanostructure region, while significant transverse heat flow is observed at the sidewalls of the nanostructures. This means that the nanostructures change the direction of heat transfer, suppressing phonon transport at the interface, but traveling new thermally conductive pathways in the sidewalls, thus transferring more heat. This phenomenon has also been observed in previous studies of simple rectangular nanostructures [20,21], which is mainly caused by multiple reflections of phonons inside the nanostructures (as shown in Figure 7(a)). The results of the temperature and heat flow distributions calculated using the LBM method are given in Figure 4(m-o), where $\beta$ is 1 as an example. Similar to the MC results, the nanostructured interfaces calculated by the LBM method also showed regionally distributed heat flow and significant lateral transport heat flow. The distribution of heat flow is more regular, limited by the size of the D2Q8 lattice used, but the overall results are consistent with those of the MC method. As a result, the nanostructured interface leads to non-uniform heat flow and travels new thermal conduction pathways in the nanostructured sidewalls, which is one of the main reasons for its enhanced heat transport.

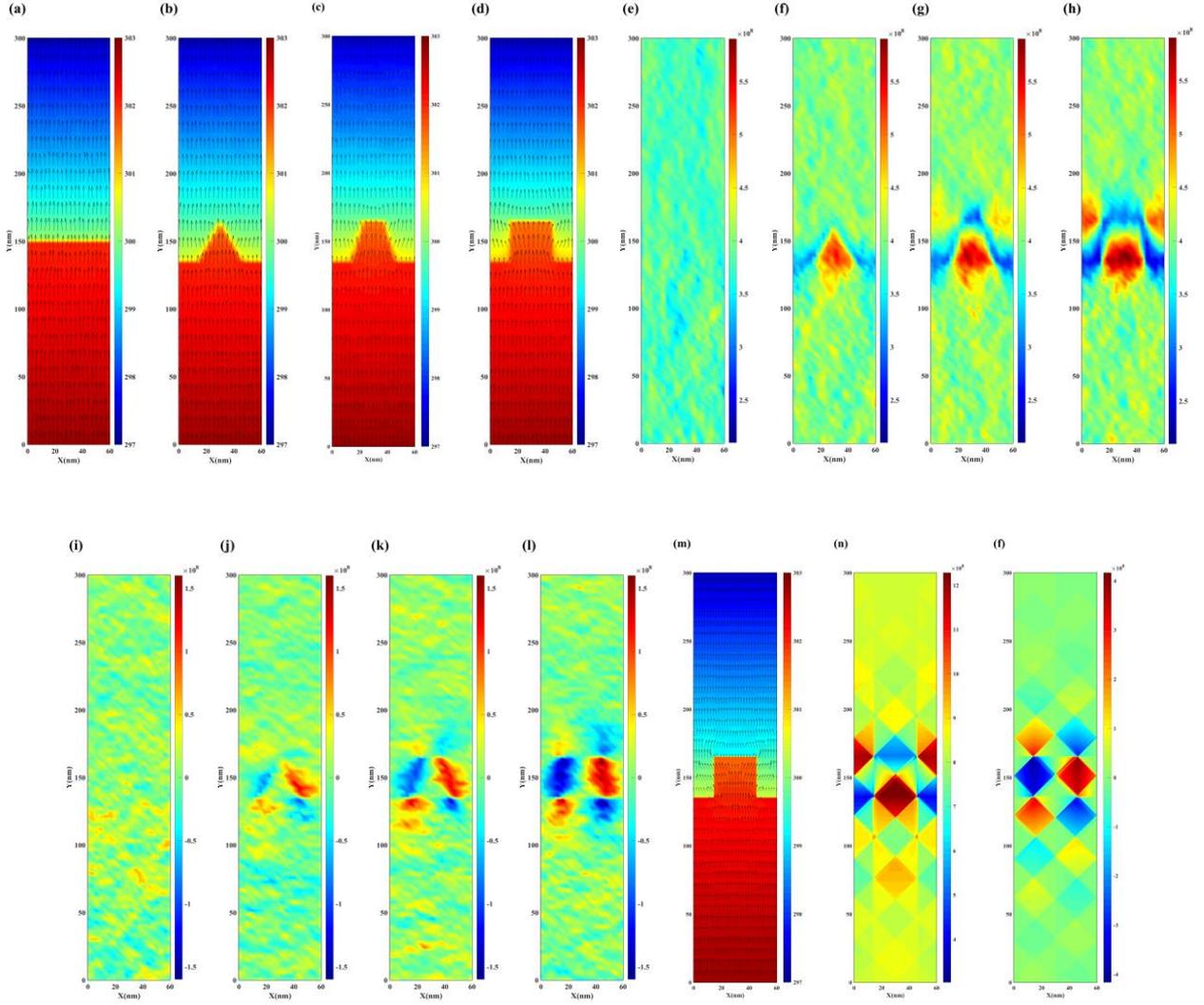

Fig. 4. Comparisons of temperature (a-d), y-direction heat flow (e-h), and x-direction heat flow (i-l) distributions of planar interface and nanostructured interfaces with $\beta$ of 0, 0.5, and 1, which are calculated by the MC method. Temperature, y-direction heat flow, and x-direction heat flow distributions (m-o) at nanostructured interfaces with $\beta$ of 1 calculated by the LBM method.

The spectral thermal conductance of the planar interface and nanostructure interfaces with different $\beta$ is provided in Figure 5(a). The spectral thermal conductance is mainly distributed in three frequency ranges, 1~4 THz, 4~6.5 THz, and 6.5~9.5 THz, and the variation of nanostructure morphology does not have a significant effect on this distribution. The peaks of the spectral thermal conductance occur near 2.7 THz, 5.2 THz, and 7.3 THz, which is because there are large phonons numbers and free paths in these frequency intervals as seen from Figure 5(b), both of which are positively correlated with the material's heat transfer properties [43]. The nanostructured interface mainly enhances the heat transferred by phonons near the frequencies of 2.7 THz and 5.2 THz compared to the planar interface. This is mainly attributed to the nanostructures increasing the probability of the reflected phonon encountering the interface again (before the next phonon-phonon scattering occurs), thus increasing the phonon transmittance. This strengthening mechanism is more effective for low-frequency phonons with large free paths. When the phonon's free path is small, it is very susceptible to phonon-phonon scattering, which transforms it into phonons of other frequencies, so the nanostructures have less effect on it. According to DMM theory, the interface transmittance is related to the angle between the phonon group velocity and the interface [40]. When $\beta$ is 0, the spectral thermal conductance of phonons near the 5.2 THz is lower than that of the planar interface because the inclined sidewalls reduce the phonon transmittance. When $\beta$ shifts from 0 to 1, the spectral thermal conductance near 2.7 THz decreases, but elevates near the 5.2 THz. According to Figure 5(c), the transmittance enhancement brought about by the nanostructures to increase the probability of phonon encountering the interface is not significant, since the low-frequency phonons near 2.7 THz already have high transmittance. For phonons near 5.2 THz, on the other hand, their transmittances are at the low values of <0.4, which implies that multiple reflections can significantly enhance the probability of transmitting through the interface, bringing about a very high spectral thermal conductance contribution. For phonons from 6.5 to 9.5 THz, they essentially do not benefit from the area expansion because of their inherently high transmittance and the small free path.

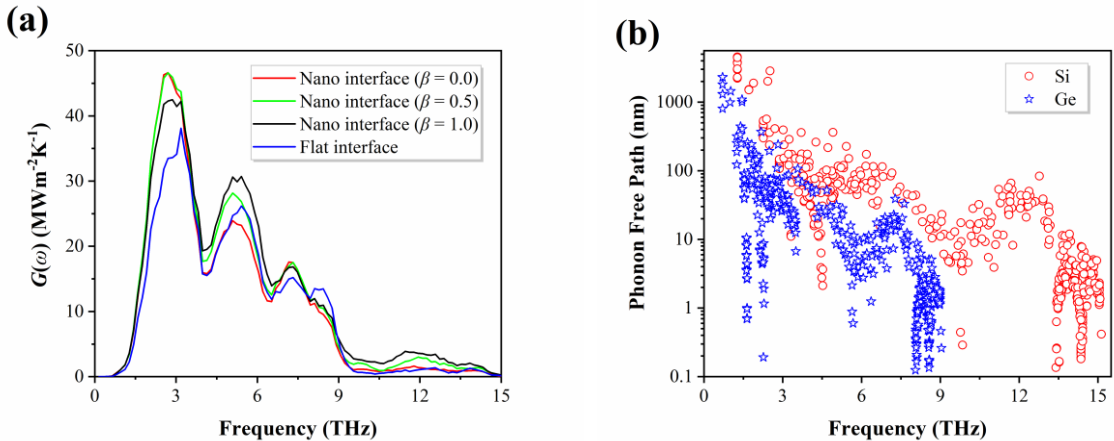

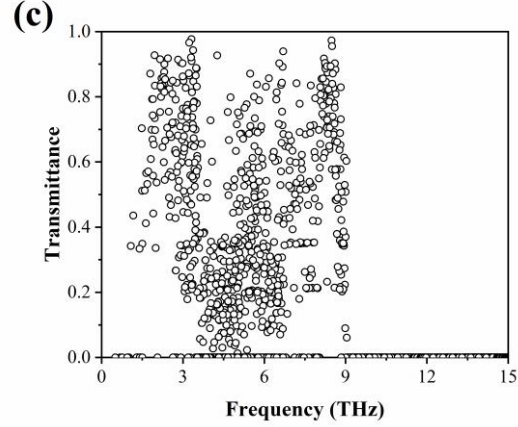

Fig. 5. (a) The spectral thermal conductance of the planar interface and nanostructure interfaces with different $\beta$. (b) Phonon free paths of Si and Ge. (c) Phonon spectrum transmittance.

## 3.3 Effect of rectangular nanostructure dimension

The ITC enhancement ratio and the average phonon scattering number for rectangular structures of different sizes are presented in Figure 6. When $W$ is different, the strengthening effect of increasing $H$ varies. When $W$ is 20 nm and 40 nm, $R_G$ first increases and then decreases when $H$ increases. The optimal structure height is 40 nm when $W$ is 20 nm and 60 nm when $W$ is 40 nm. When the nanostructures have a large width, $R_G$ increases gradually as the $H$ grows, but the increment rate decreases. For the currently studied rectangular nanostructure dimensions, the maximum $R_G$ is 1.31, and further increasing the nanostructure height can bring about greater enhancements. When the height of the structure is low, a smaller $W$ is more beneficial to enhance the interfacial heat transfer because, according to Figure 6(b), this obtains a larger $N_{ave}$. However, as the $H$ increases, there will exist an optimal $W$ that maximizes $R_G$. The reason is that the small $W$ significantly increases the probability of phonons hitting the interface when $H$ is fixed. After phonons transmit through the interface, they may not undergo phonon-phonon scattering and transfer heat, but directly encounter the interface and transmit back into the original material (double transmission phenomenon), as shown in Figure 7(b). This double transmission phenomenon leads to the possibility that the two phonon-interface scattering processes may not have realized heat transfer across the interface materials. In addition, a large $W$ reduces the probability of a phonon encountering the interface, thus there exists an optimal $W$ enabling the largest $R_G$ at a fixed $H$. The optimal $W$ is 40 nm when $H$ is 40 nm and 80 nm when $H$ is 80 nm. Therefore, when $H$ is determined, using $W$, which is close to $H$, is a good choice.

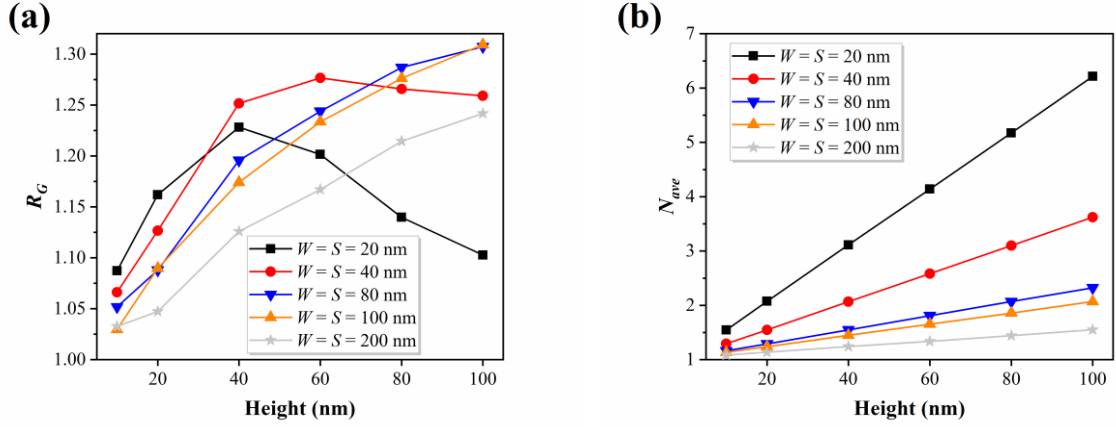

Fig. 6. The interface thermal conductance enhancement ratio (a) and the average phonon scattering number (b) for rectangular structures of different sizes.

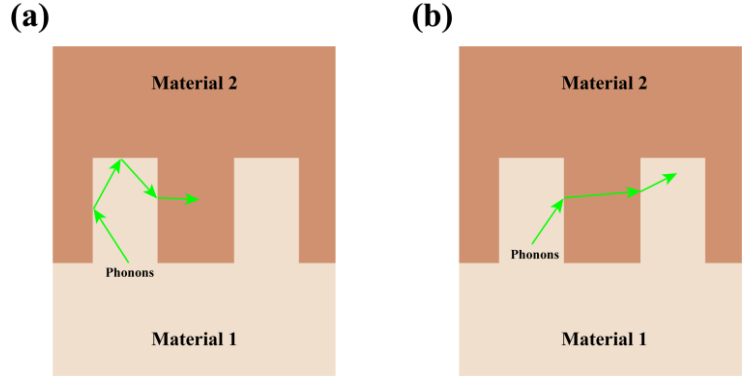

Fig. 7. Phonon-interface scattering phenomena at nanostructured interfaces. (a) Multiple reflections (b) Double transmission.

Figure 8 (a-f) show the y-direction and x-direction heat flow distributions for nanostructures with $W$ of 20 nm and $H$ of 20, 40, and 60 nm. As the height of the nanostructure increases, the heat accumulation is more significant, which means that more heat needs to be transferred through the sidewalls. The reason is that the deeper square-cavity structure significantly increases the phonon scattering between the sidewalls, with elevated heat transfer from the sidewalls but reduced heat transfer by the horizontal contact surfaces. Comparing Figure 8 (b, d, and f), it is found that the transverse heat flux of the nanostructure's sidewalls gradually decreases with the increase in height, which can be attributed to the double transmission phenomenon. Figure 8 (g-l) illustrate the y-direction and x-direction heat flow distributions for nanostructures with $W$ of 100 nm and $H$ of 20, 40, and 60 nm. Similarly, increasing the height results in greater aggregation of y-direction heat flow and an increase in the extent of non-uniform heat flow. However, unlike the results for $W$ of 20 nm, increasing the height also augments the x-direction heat flow for $W$ of 100 nm. This is because when $W$ is large, the transmitted phonons do not easily return to the raw material but directly undergo phonon-phonon scattering, which weakens the effect of the double transmission phenomenon. Therefore, the large transverse heat flow at the nanostructure sidewalls is the key to achieving efficient interfacial heat transport, which is strongly related to the sidewall height and the spacing between the sidewalls.

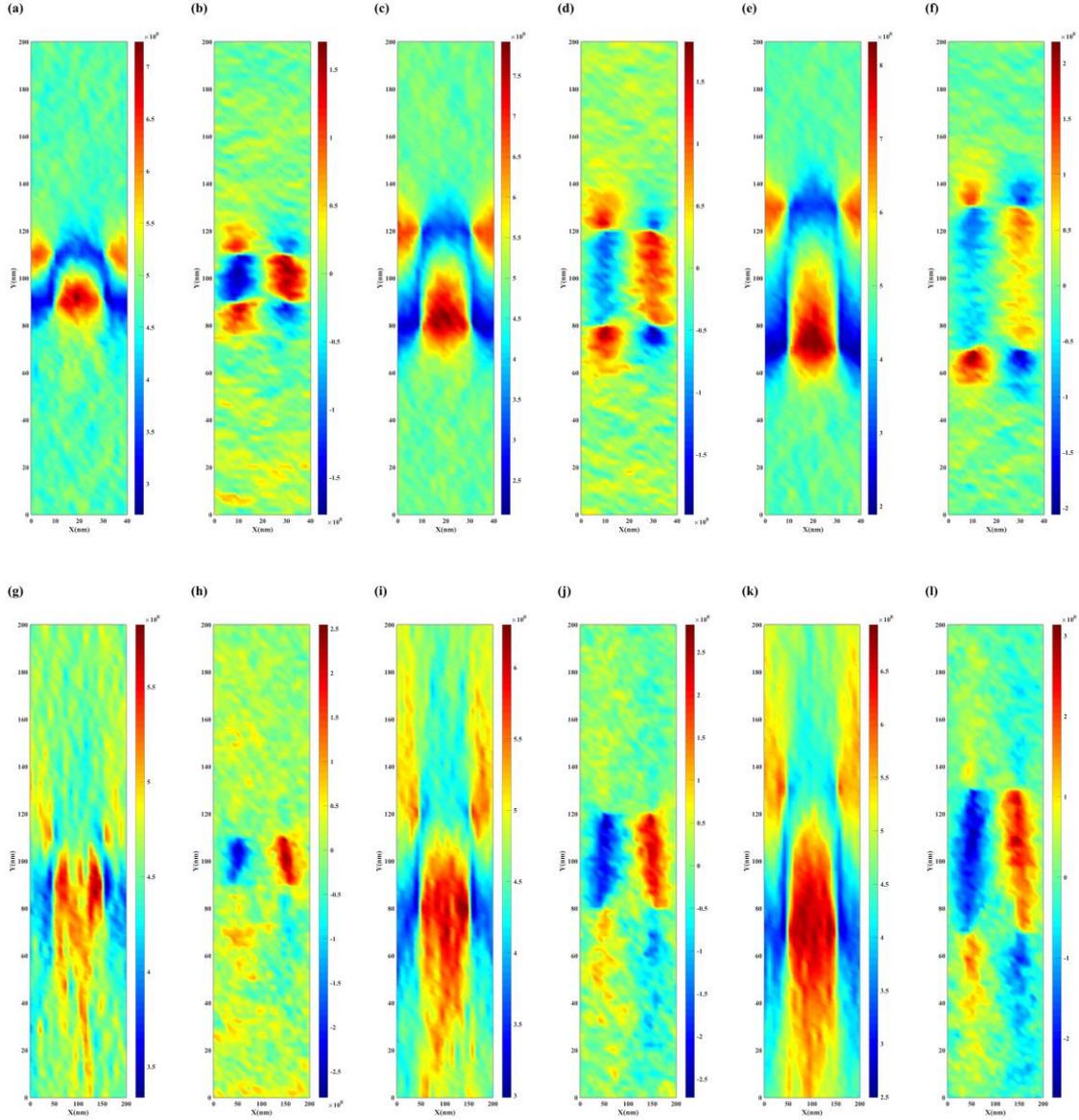

Fig. 8. The y-direction and x-direction heat flow distributions for nanostructures with (a) *W* of 20 nm and *H* of 20 nm (b) *W* of 20 nm and *H* of 40 nm (c) *W* of 20 nm and *H* of 60 nm (d) *W* of 100 nm and *H* of 20 nm (e) *W* of 100 nm and *H* of 40 nm (f) *W* of 100 nm and *H* of 60 nm.

Figure 9 demonstrates the spectral thermal conductance of rectangular nanostructured interfaces with different sizes. As can be seen from Figure 9 (a) and (b), a larger *H* is more unfavorable for heat transfer by phonons with frequencies from 6.5 to 9.5 THz when *W* is fixed. The reason is that the large interfacial transmittance and small free path of phonons with these frequencies make them not benefit from the enhanced phonons multiple reflections phenomenon caused by the increase of *H*. When W is 20 nm and H is 40 nm, the spectral thermal conductance of the phonons with frequencies from 1 to 4 THz and from 4 to 6.5 THz is maximum. As *H* continues to increase, there is a large decrease in the spectral thermal conductance of phonons with frequencies from 1 to 4 THz, while a small decrease in the spectral thermal conductance of phonons with frequencies from 4 to 6.5 THz is observed. This is because for phonons with smaller frequencies, which have larger free paths, the double transmission phenomena that weaken phonon transmission is more likely to occur. According to Figure 9(c) and (d), changing *W* has almost no effect on the spectral thermal conductance of phonons with frequencies greater than 6.5 THz when *H* is fixed. For phonons with frequencies from 1 to 4 THz and 4-6.5 THz, the spectral thermal conductance first increases and then decreases as the *W* rises. The spectral thermal conductance of phonons with frequencies from 1 to 4 THz varies more because these phonons have larger free paths and are more sensitive to structure variation. When *W* is small, it favors the phonons' multiple reflections, but also increases the possibility of the double transmission; when *W* is large, it does not favor the multiple reflections, but the possibility of the double transmission drops. Thus, the phenomena of multiple reflection and double transmission of phonons together determine the enhancement effect of nanostructures on interfacial heat transfer.

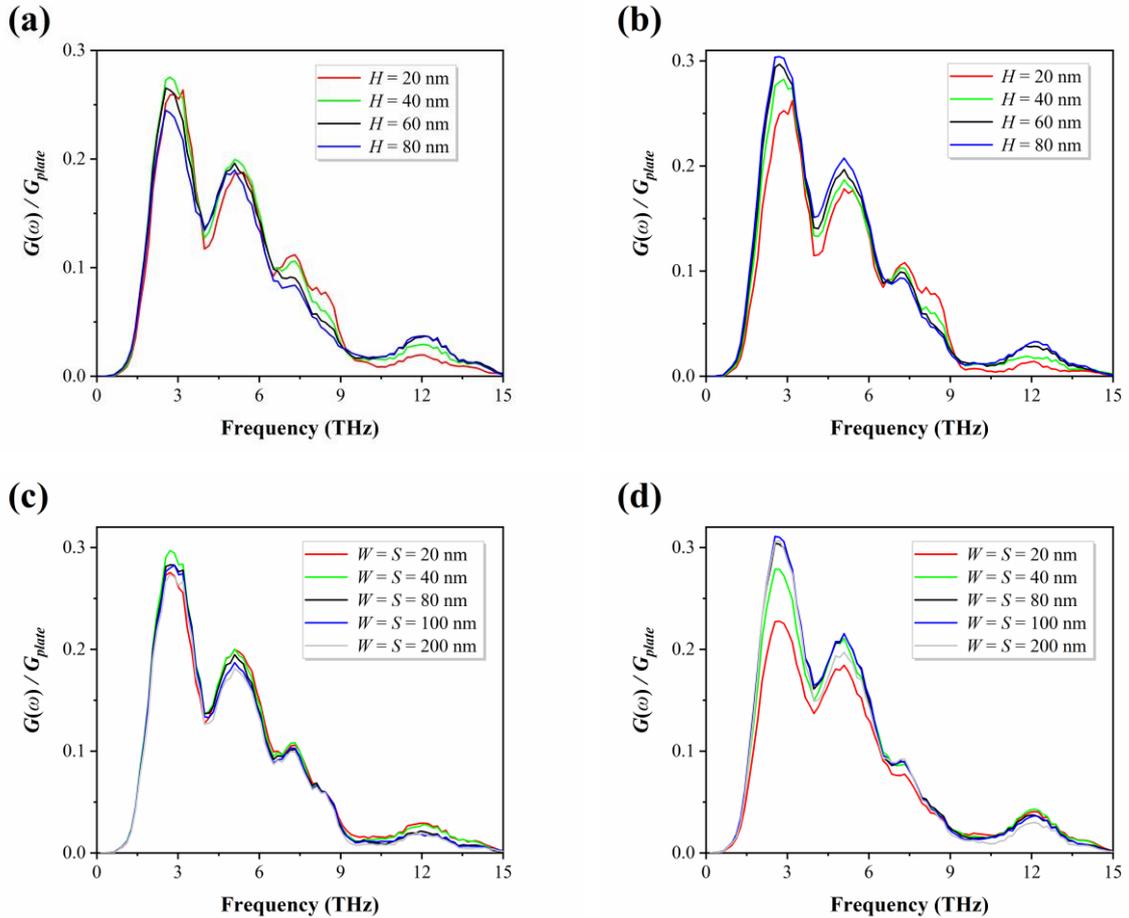

Fig. 9. The spectral thermal conductance of rectangular nanostructured interfaces when (a) *W* is 20 nm (b) *W* is 100 nm (c) *H* is 20 nm (d) *H* is 100 nm.

## 4 Conclusions

In this paper, a multiscale method to study the thermal transfer at nanostructured interfaces is developed by combining density functional calculation, Monte Carlo simulation, and diffuse mismatch method. The changes in the transport paths and contributions to thermal conductance of different frequency phonons caused by changes in nanostructure morphology and size are investigated. The main conclusions are as follows:

(1) Compared to the triangular and trapezoidal nanostructures, the rectangular nanostructures are more beneficial in enhancing the probability of the reflected phonons encountering the interface, and thus the phonon interfacial transmittance. The multiple reflections are more favorable to enhance the interfacial transport of phonons with larger free paths but low transmittance, while the interfacial transport enhancement brought by nanostructures is not significant for phonons with small free paths or high transmittance.

(2) The nanostructure makes the interfacial heat flow extremely heterogeneous, with significant transverse heat flow occurring at the sidewalls, resulting in specific thermal conduction pathways. When the spacing between the nanostructure's sidewalls is small, the transmitted phonons may reencounter the interface and transmit back into the original material. This double transmission phenomenon weakens the transverse heat flow through the nanostructure's sidewalls and has a significant effect on low-frequency phonons with large free paths.

(3) The phenomena of multiple reflections and double transmission together result in the existence of the optimal nanostructure dimension that maximizes the nanostructures enhancement effect on interfacial heat transfer. The optimal nanostructure width is 100 nm when the height is 100 nm and the maximum interfacial thermal conductance enhancement ratio is 1.31.

These results can provide guidance for the design of heat transfer enhancement structures at the interface of the actual high-power chips.

**Acknowledgments**

This work was supported by the National Natural Science Foundation of China (Grant NO. U20A20299, 52006102), the Fundamental Research Funds for the Central Universities (No.30923010917).